\documentclass[11pt]{article}
\usepackage[T2A]{fontenc}
\usepackage[cp1251]{inputenc}
\usepackage[english]{babel}
\usepackage{graphicx}
\usepackage{booktabs}
\usepackage{amsmath}
\usepackage{amssymb}
\usepackage{ragged2e}

\newcommand{\prtwo}{PR~2$^{\circ}$}
\newcommand{\prfour}{PR~4$^{\circ}$}
\newcommand{\prsix}{PR~6$^{\circ}$}

\long\def\maintitle#1{{\vskip 20mm \begin{center}\section*{#1}\end{center}\nopagebreak[4]}}

\long\def\author#1{{\begin{center}\normalsize{\bf#1}\end{center}\vskip-1em\index{#1}}\nopagebreak[4]}
\long\def\address#1{{\begin{center}\small\noindent#1\end{center}\vskip-8mm}\nopagebreak[4]}

\begin{document}
\noindent\mbox{\small The 13$^{th}$ EVN Symposium \& Users Meeting Proceedings, 2016}

\maintitle{MULTIVIEW PHASE CORRECTIONS AT LOW FREQUENCIES FOR PRECISE ASTROMETRY}

\author{G.~Orosz$^{1}$, M.~J.~Rioja$^{2,3,4}$, R.~Dodson$^{2}$, H.~Imai$^{1}$, and S.~Frey$^{5}$}

\address{$^{1}$Graduate School of Science and Engineering, Kagoshima University, Japan\\$^{2}$International Centre for Radio Astronomy Research, Australia\\$^{3}$CSIRO Astronomy and Space Science, Australia\\$^{4}$Observatorio Astron\'{o}mico Nacional (IGN), Spain\\$^{5}$Konkoly Obs., MTA Research Centre for Astronomy and Earth Sci., Hungary}

\begin{abstract} We present a multi-calibrator solution, i.e. MultiView, to achieve accurate astrometry on the level of the thermal noise at low VLBI frequencies dominated by ionospheric residuals. We demonstrate on L-band VLBA observations how MultiView provides superior astrometry to conventional phase referencing techniques \cite{rioja2017}. We also introduce a new trial method to detect antenna based systematic errors in the observations \cite{orosz2017}. All presented methods and results are based on our recent papers \cite{orosz2017,rioja2017}.
\end{abstract}

\section{Introduction} \label{introduction}

High-precision astrometry using very long baseline interferometry (VLBI) depends on the precise calibration of fringe phase residuals. Difficulties usually arise from residual errors in the propagation medium that were not addressed properly with the adopted phase referencing (PR) method. Conventional PR techniques at roughly 8--43\,GHz are effective in achieving astrometric accuracies on a 10 $\mu$as-level (\cite{reid2014}, and references therein), but the short tropospheric coherence times at higher frequencies and the spatial ionospheric irregularities at lower frequencies degrade the effectiveness of existing solutions. Recently, new techniques that rely on (near) simultaneous multi-frequency observations have extended the limit of $\mu$as astrometry to the mm-wavelength regime (see Frequency Phase Transfer and Source Frequency PR, e.g. \cite{rioja2015}).

At frequencies below $\sim$8\,GHz the dispersive ionospheric propagation effects have an increasingly dominant signature. The slowly changing spatial irregularities of plasma density in the atmosphere introduce differential path variations between the direction of the calibrator and target sources and cause systematic position errors even for small source separations. Existing solutions include using wide-spread bands to measure and remove the frequency-dependent dispersive component of ionospheric delays (see \cite{brisken2000,brisken2002} or the newly developed Multi-Frequency PR \cite{dodson2016}); or use simultaneous observations of a target--calibrator pair in the same primary beam of the VLBI antennas (in-beam PR) to minimize angular separation and thus any residual errors (see e.g. \cite{deller2016}). However, wide bands can only be used for continuum sources and in-beam PR is limited by the availability of nearby calibrators.

An alternative approach is to use scans on multiple calibrators to model the 2D phase screen around the target, a technique termed as MultiView (MV) calibration \cite{rioja2002,rioja2009,dodson2013}. By determining the spatial structure of the ionosphere, it is possible to reach astrometric errors only limited by the random thermal noise, even with calibrators several degrees away. This increases the availability of suitable calibrators and extends the applicability of precise low-frequency astrometry. In this contribution we shortly introduce an observational demonstration of MV for astrometry at 1.6\,GHz. The major part is a summary of our results published in \cite{rioja2017} and we refer the reader to this publication for a complete discussion on the presented topics.

\section{Observations and data reduction} \label{observations}

Observations were conducted at 1.6\,GHz in two epochs with the NRAO Very Long Baseline Array (VLBA) separated by one month (ID: BO047A7 and A4). Both sessions included observations of an OH maser source (WX~Psc) and a continuum source (J0106+1300) in the same primary beam (separated by 24$^\prime$), and three other calibrators spread around the targets with increasing separations (J0113+1324 at 2$^{\circ}$, J0121+1149 at 4$^{\circ}$ and J0042+1009 at 6$^{\circ}$) selected from the Astrogeo Center Database. The sources were observed with 5-min duty cycles for 4 hours per epoch, as longer source-switching times from earlier observations proved too long to reliably connect the phases between scans. Scans were recorded at a data rate of 256\,Mbps, using 4 IF bands spread out over 300\,MHz including all ground-state OH maser lines, each with bandwidths and channel spacings of 8\,MHz and 1.95\,kHz, respectively.

The goal of the observations were to demonstrate MV by conducting a comparative astrometric study using various calibration techniques. These include conventional PR techniques between J0106+1300 as target and J0113+1324 (\prtwo), J0121+1149 (\prfour) or J0042+1009 (\prsix) as calibrators; in-beam PR between WX~Psc as target and J0106+1300; and MV calibration techniques using solutions based on J0113+1324, J0121+1149 and J0042+1009 to apply to J0106+1300 or WX~Psc as targets to test MV on both continuum and line datasets. The PR analyses were carried out using standard procedures in the NRAO Astronomical Image Processing System (AIPS), while the MV analysis required additional steps to incorporate direction dependent effects in the calibration. This was done by a 2D interpolation -- including the tracking of possible 2$\pi$ phase ambiguities -- of fringe phase solutions along the directions of all calibrators to provide corrections along the line of sight of the target observations, thus solving for the first order effects of the spatial structures in the propagation medium above each antenna. As a result, the target was phase referenced to an assumed virtual point on the sky that was derived from the combination of the three MV calibrators.

\section{Results and error analysis} \label{results}

\begin{figure}[t]
\centering
\includegraphics[width=1\textwidth]{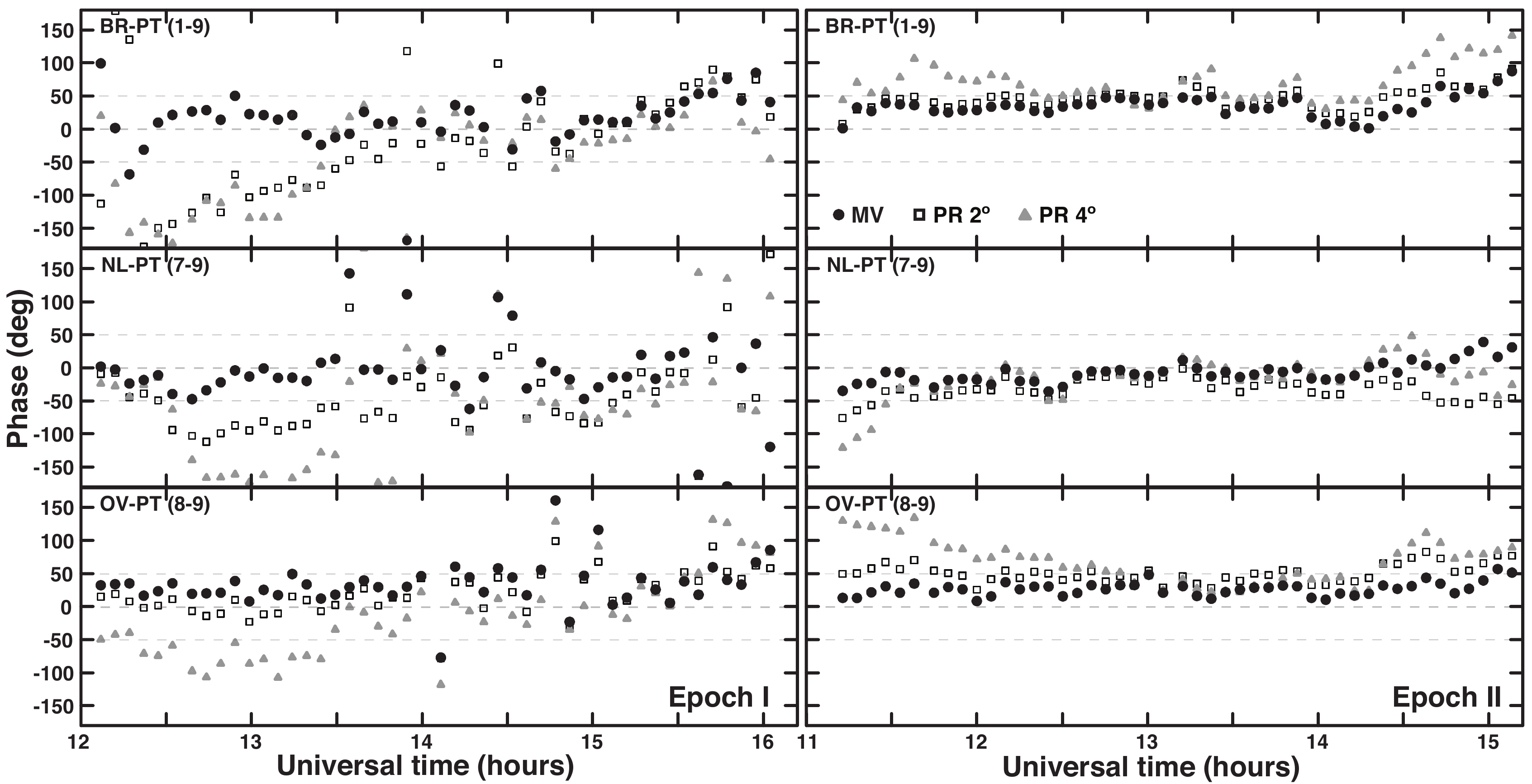}
\caption{\label{fig:vis}Calibrated visibility phases of J0106+1300 using MV (solid circles) with three calibrators, and PR with a calibrator 2$^{\circ}$ (squares) and 4$^{\circ}$ (triangles) away on a subgroup of baselines, in Epoch I (left) and Epoch II (right).}
\end{figure}

We obtain MV solutions for the quasar J0106+1300 and the maser WX~Psc, using the same two observing sessions and calibrators, which we then compare to several traditional PR solutions. Figure~\ref{fig:vis} shows the residual visibility phases of J0106+1300 after calibration using \prtwo, \prfour and MV. Slow phase drifts after conventional PR indicate residual systematic phase errors, which are different in the two epochs and depend on the ionospheric weather conditions and the angular separation of the calibrator. The large disturbances seen at the beginning of the observations correspond to sunrise. The MV phase residuals are the smallest in all cases and show no systematic trends. They are also similar in both epochs, indicating the effective mitigation of ionospheric phase errors regardless of weather conditions (see more in \cite{rioja2017}).

\begin{figure}[htb!]
\centering
\includegraphics[width=1\textwidth, trim=2mm 7.8mm 30mm 2mm, clip]{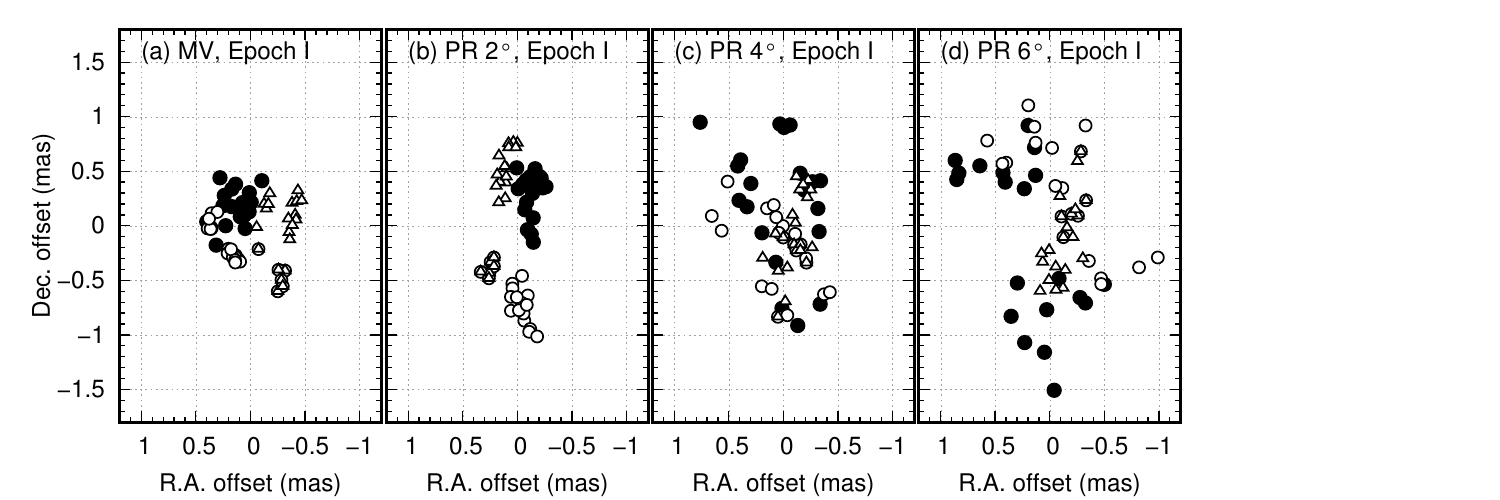}
\includegraphics[width=1\textwidth, trim=2mm 0mm 30mm 2.8mm, clip]{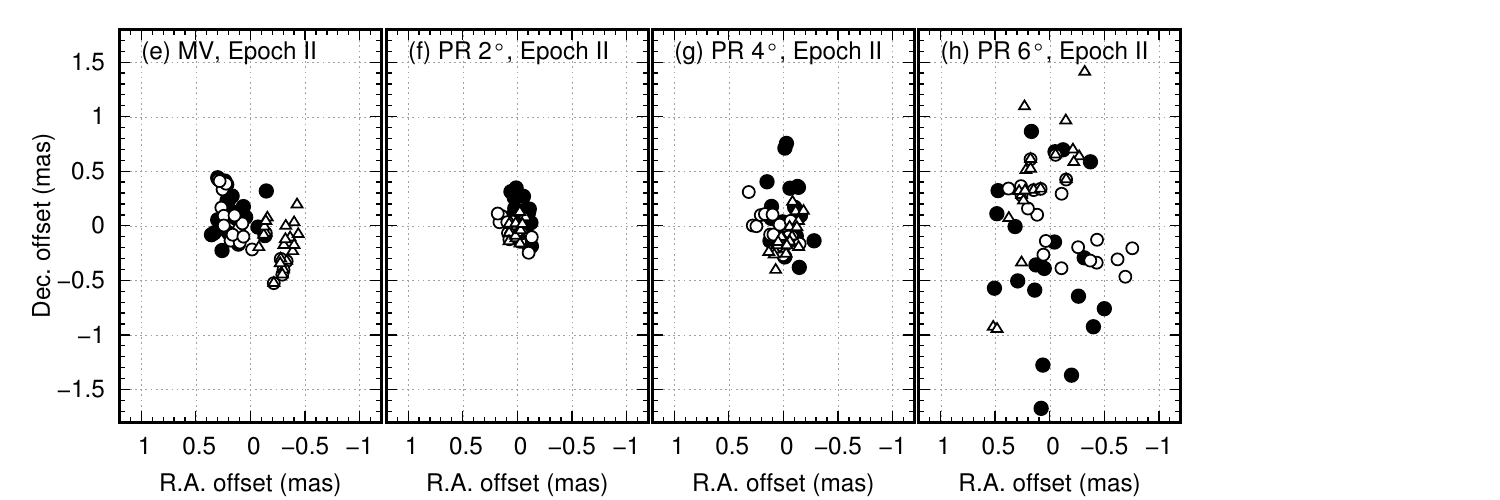}
\caption{\label{fig:tri}Positions of J0106+1300 in respective epochs using the various solutions, determined with all the 56 three-antenna subarrays of the continental VLBA (not incuding MK or SC). Open circles and triangles mark measurements that contain antennas BR and HN respectively. Offsets are relative to the mean positions.}
\end{figure}

Residuals are composed of systematic and random errors. Systematic effects are hard to detect as they introduce shifts in the position without degrading the resulting image quality. We introduce a new technique to try and separate these two types of error sources, by imaging the calibrated visibilities with all possible three-antenna subarrays of the VLBA \cite{orosz2017}. We then measure the peak positions in the resulting maps, which are all coherent but shifted images of the same source. Comparing the positions determined from these subarrays can expose antennas affected by systematic errors. Figure~\ref{fig:tri} demonstrates the results from subarray imaging for J0106+1300, using all possible PR and MV calibrations. The spread of the derived positions is related to random errors and it gets larger with increasing calibrator separations. In order to show how antennas can be related to shifts depending on calibration method and ionospheric conditions, BR (circles) and HN (triangles) were highlighted as being the most remote antennas in the continental VLBA (MK and SC had to be flagged out earlier). In \prsix and \prfour we do not see any particular pattern, but systematic shifts become visible in the \prtwo and MV solutions due to the smaller scatter. Using the methods described in \cite{asaki2007} and \cite{orosz2017}, we expect residual ionospheric errors to be on the order of $\sim$0.1\,mas for MV and $\sim$0.6\,mas for \prtwo, which are comparable to the shifts seen for BR in Epoch I and demonstrate the additional corrections in MV. However, the MV solutions also show a small peculiar offset for HN, almost identical in size and direction in both epochs, that is also visible in Epoch I of \prtwo although in the opposite direction. Understanding this effect requires further study. For more details on sub-array imaging and analysis of the maser data, refer to \cite{orosz2017}.

Table~\ref{table:errors} shows the total astrometric uncertainties of our measurements for J0106+1300 using various estimates. Thermal noise errors are derived from the random thermal noise in the maps and are not sensitive to systematic effects. We also estimate the errors by fitting a Gaussian model to the distribution of peak positions shown in Fig.~\ref{fig:tri}, and are termed as ``triangle baseline errors'' that should give more conservative estimates. Finally, we calculate the repeatability errors which are derived from the change in the measured offsets at the two epochs and provide estimates on the total astrometric accuracy including all remaining systematic errors. Only MV is capable of reproducing the position of J0106+1300 on the level of the calculated uncertainties in each epoch and it provides more than an order of magnitude increase in accuracy compared to \prtwo. For a detailed discussion on these and the maser results, refer to \cite{rioja2017}.

\begin{table}[htb!]
\centering
\caption[]{\label{table:errors}Astrometric error estimates for J0106+1300.}
\begin{tabular}{l c c c c c}
\hline
\hline
\noalign{\smallskip}
Method & \multicolumn{2}{c}{Thermal (mas)} & \multicolumn{2}{c}{Triangle (mas)} & \multicolumn{1}{c}{Repeatability (mas)}\\
\noalign{\smallskip}
& I & II & I & II & $\Delta'_{I-II}$ \\
\noalign{\smallskip}
\hline
\noalign{\smallskip}
MV      & 0.17  & 0.14   & 0.36   & 0.33  & 0.10  \\
\prtwo  & 0.19  & 0.11   & 0.56  & 0.14  & 2.72   \\
\prfour & 0.42  & 0.22   & 0.54  & 0.26  & 3.94	   \\
\prsix   & 0.75  & 0.44   & 0.70  & 0.70  & 11.2   \\
\noalign{\smallskip}
\hline
\end{tabular}\\
\vspace{-10 pt}
\justify
{\bf Notes.} Thermal: errors from thermal noise. Triangle: triangle baseline errors.
\end{table}

\section{Conclusions} \label{conclusions}

We demonstrated the superior calibration results of MV compared to conventional PR techniques, by achieving accurate astrometry on the level of the thermal noise ($\sim$0.1 mas). These pilot results indicate that the multi-calibrator phase-modeling approach of MV can be the key to complete ionospheric mitigation and $\mu$as astrometry at low frequencies. This is especially important with the arrival of the Square Kilometer Array (SKA), as SKA will operate in the ionosphere-dominated frequency regime and is expected to have multi-beam capabilities ideal for MV \cite{paragi2015}. Multiple beams can make MV calibration even more accurate as they remove dynamic errors and coherence time limitations inherent in any source-switching setup. Multi-beam capabilities are already available for a couple of antennas, e.g. ASKAP, WSRT Apertif, Effelsberg in the 800--1800 MHz range or the prototype system at the Sardinia Radio telescope, currently operating with 5 beams in C-band but expected to expand both in frequency and number of beams.

\subsection*{Acknowledgements}
GO, MR, RD and HI acknowledge support from the DFAT grant AJF-124. GO and HI have been supported by the JSPS Bilateral Collaboration Program and KAKENHI programs 25610043 and 16H02167. GO had support from Monbukagakusho:MEXT, the EVN Symposium SOC and Kagoshima University.

\end{document}